\def\beq{\begin{equation}}
\def\eeq{\end{equation}} 
\def\beqn{\begin{eqnarray}}
\def\eeqn{\end{eqnarray}}

\newcommand{\etal}{{\it et al}}

\def\AEF{A.E. Faraggi}

\def\MODA{{\em Mod. Phys. Lett.} A}
\def\IJMP{{\em Int. J. Mod. Phys.} A}

\input epsf

\documentclass{Rinton-P9x6}

\begin{document}

\title{Phenomenological Survey of M--theory
\footnote{Invited talk presented at SUGRA 20, Boston MA, March 17--21 2003.}
\footnote{OUTP--03--17P}
}
\author{Alon E. Faraggi}

\address{Theoretical Physics Department, University of Oxford,
Oxford OX1 3NP, UK
\\E-mail: faraggi@thphys.ox.ac.uk}


\maketitle

\abstracts{The Standard Model data suggests that the quantum gravity vacuum
should accommodate two pivotal ingredients. The existence of 
three chiral generations and their embedding in chiral 16 SO(10)
representations. The $Z_2\times Z_2$ orbifolds are examples
of perturbative heterotic string vacua that yield these properties.
The exploration of these models in the nonperturbative
framework of M--theory is discussed. A common prediction of these
constructions is the existence of super--heavy meta--stable states
due to the Wilson--line breaking of the GUT symmetries.
Cosmic ray experiments in the forthcoming years offer
an exciting experimental window to the phenomenology of such
states.}

\section{Introduction}
\begin{figure}[t]
\centerline{\epsfxsize 3.0 truein \epsfbox{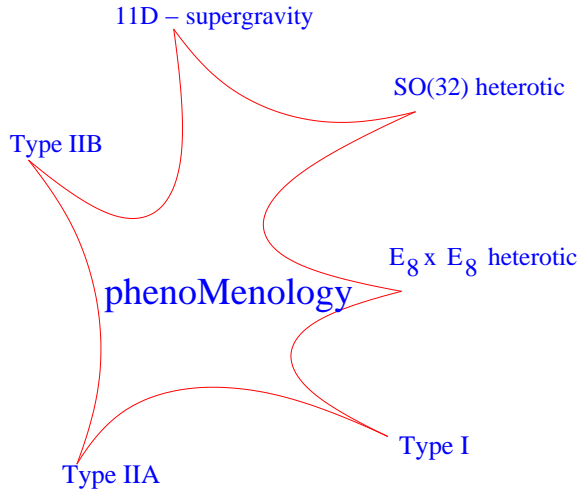}}
\caption{M--theory picture of string theory}  
\label{pati0405022}
\end{figure}

Over the past few years substantial progress has been achieved
in the basic understanding of string theory. The picture which emerged, 
and which is depicted qualitatively in figure \ref{pati0405022}, is
that the different string theories in ten dimensions are limits
of a more fundamental theory, traditionally dubbed M--theory.
The question remains how to connect these advances, 
and string theory in general, to experimental data. 
I think that it will be universally agreed that this
is a vital question that string theory faces, and
opinions may differ on what is the most suitable methodology to
advance this issue. 

In this respect I think that there are some prevailing misconceptions.
The first question that should be posed is why one
should be interested in string theory in the first place.
Physics is first and foremost an
experimental science, and after all the recent celebrated
advances in string theory have to do with unification
of theories in ten or eleven dimensions, and what has this
to do with experimental physics? Nevertheless,
given the present experimental data, the exploration of string theory
is well motivated.

One misconception is the reference
to string theory as the ``theory of everything''. I think that 
it is besides the point. The primary questions in my view are 
those of relevance and utility. Namely, the experimental data
that we observe reveal patterns of gauge charges and mass spectra.
The main issue is whether the structures that appear in string
theory are relevant and can be utilized toward understanding
the physical origin of the observed patterns. In this respect
another misconception is the view that the sole merit of
string phenomenology is to find the one true vacuum that
corresponds to the observed world. While this is a well 
posed goal that is pursued with vigor and intent, I think that
it is again besides the point. In the first place,
as is especially evident
following the string duality developments, none of
the string limits can fully characterize the true vacuum.
The true vacuum should have some nonperturbative realization.
The perturbative string theory limits and the eleven dimensional classical
limit are effective limits that can at best probe some of the properties
of the true vacuum. In this view it may well be that some limits
may highlight some properties of the vacuum, whereas other
limits may be more instrumental to extract different properties.
A good example of this is the dilaton stabilization problem. 
As is well known, in the perturbative heterotic--string limit
the dilaton, whose VEV governs the string gauge and gravitational
couplings, has a run--away potential and cannot
be stabilized at a finite value. However, we should regard 
the heterotic limit as the zero coupling expansion of the more
basic theory. With our 
present understanding of string theories in the context
of their M--theory embedding it is clear that we should not
in fact expect the dilaton to be stabilized in the heterotic limit. 
In order to stabilize the dilaton we have to move away from
the zero coupling expansion, or to move away from the 
perturbative heterotic--string limit. The existence
of the classical eleven dimensional limit in which the dilaton
is interpreted as the moduli of the eleventh dimension
lends credence to this general expectation. Thus, the issue
of dilaton stabilization may be more accessible, even if not yet
fully resolved, in other limits of the underlying theory,
rather than in the perturbative heterotic string limit.
The problem of supersymmetry breaking may be similar.

\begin{figure}[t]
\centerline{\epsfxsize 3.0 truein \epsfbox{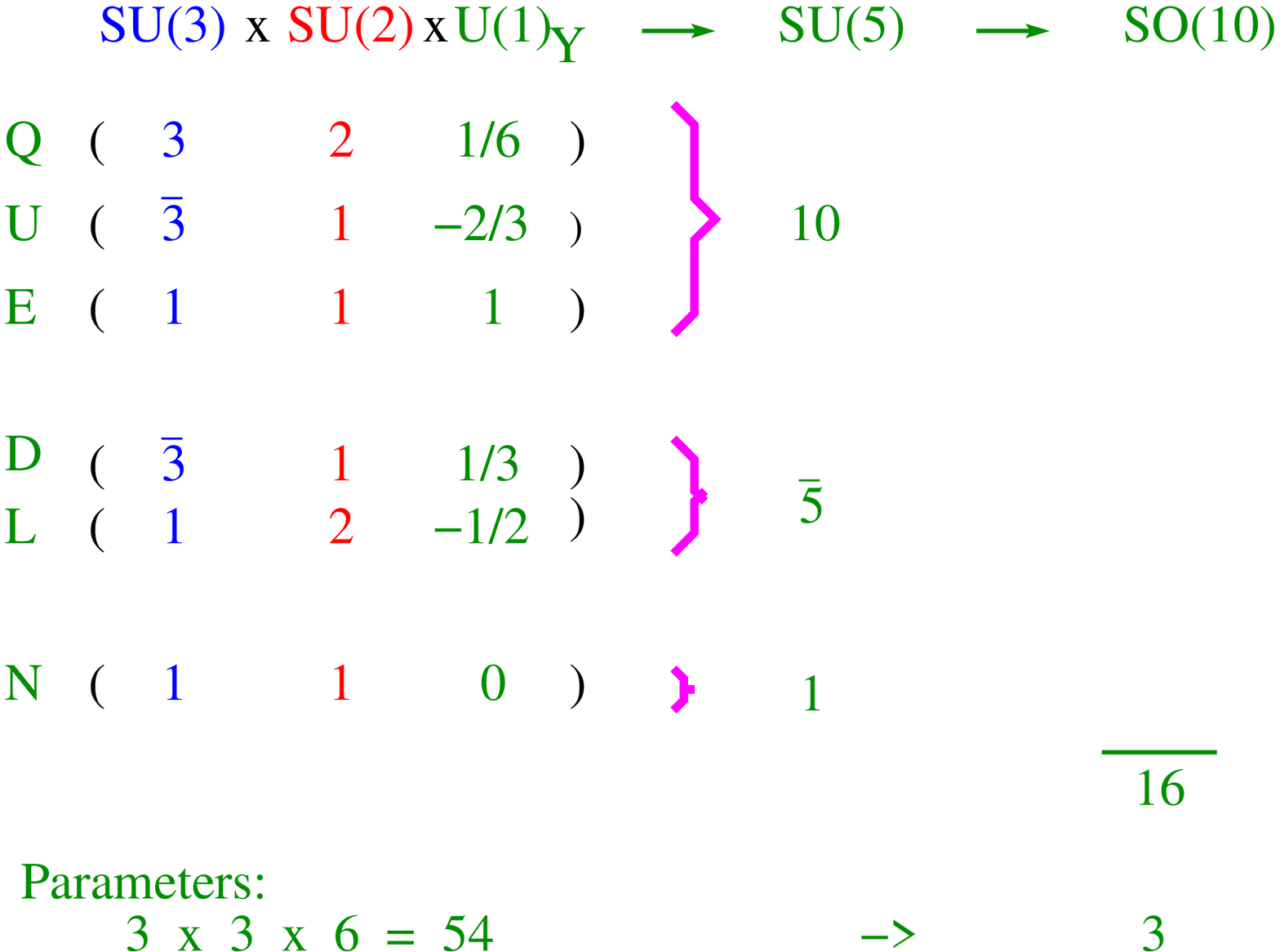}}
\caption{GUT--embedding of the Standard Model}  
\label{smmultiplets}
\end{figure}

The primary questions in respect to string and M--theories are
therefore those of relevance and utility. The relevance follows
from the basic structure of the Standard Model of particle physics. 
The Standard PArticle Model (SPAM) matter sector is composed of three chiral 
generations, charged under the three group factors $SU(3)\times
SU(2)\times U(1)_Y$. The most remarkable aspect of the SPAM, and
for me its essence, is the GUT embedding of the Standard Model
representations. Most striking is the embedding in $SO(10)$
in which each generation
fits into a single 16 spinorial representation of $SO(10)$.
The GUT embedding of the Standard Model spectrum is depicted
in figure \ref{smmultiplets}. If we regard the gauge charges
of the Standard Model states as experimental observables,
as they were in the process of its experimental discovery, then 54
parameters are needed at the level of the Standard Model to
account for the matter charges. The embedding in $SO(10)$ 
reduces this number of parameters to three, which is the 
number of chiral 16 $SO(10)$ representations needed to
accommodate the Standard Model states. 

The GUT embedding of the Standard Model is also supported by the logarithmic
running of the Standard Model parameters. Quantum field theories,
in general, and their specific realization in the form of the 
Standard Model, predict that the gauge and matter couplings
evolve logarithmically with the energy scale. This evolution
has been confirmed experimentally in the energy range accessible to
collider experiments. It is also consistent qualitatively
with the big desert scenario,
suggested by grand unified theories, in the gauge and matter sectors
of the Standard Model. The scalar sector of the Standard Model
is not protected against radiative corrections from higher
scales, and therefore the Standard Model needs to be augmented
with an additional sector that protects the scalar states 
from higher scale corrections. It is natural
to demand that the new sector preserves the successes of the 
Standard Model and its GUT embedding. Such a new sector
is provided by supersymmetry. While the jury is still out 
on the validity of supersymmetry and its
contemporary interpretation, it seems
to me as the best of all evils.

While the Standard Model, GUTs, supersymmetry and point quantum 
field theories have enjoyed impressive successes, they still fall 
short of providing a comprehensive framework for the fundamental
forces and matter that are observed experimentally. In the first
place gravity is not yet included as a quantum theory. It is gravely
unsatisfactory to have two fundamental theories, gravity and quantum
mechanics, that
are incompatible.
More concretely, from the point
of view of the Standard Model data, GUTs and supersymmetry
do not explain the origin of many of the parameters that we
observe in the Standard Model. Specifically, the existence
of flavor with its intricate mass and mixing spectrum is unaccounted
for. It is therefore plausible that the origin of these
additional structures must be sought in a theory that unifies 
gravity and the gauge interactions. 

Superstring theories are unique in the sense that they
provide exactly that. Namely, string theories give rise
to precisely the structures that are observed in the Standard 
Model, like matter and gauge spectrum, and they provide 
consistent framework for perturbative quantum gravity. 
Hence the utility of string theory. In regard to its relevance,
the jury is of course still out on that, but we note that string
theory gives rise to additional sectors, that may be precisely
what is needed to understand the detailed spectrum of the Standard Model.
These new sectors include the requirement of compactified manifolds
that may account for the existence of flavor. Thus, string theories
gives rise to the structures that may eventually prove relevant for
the understanding of the Standard Model data. Hence its relevance. 

Getting back to the qualitative picture of figure \ref{pati0405022},
the question is: how should we utilize the new understanding 
of string theories to advance its phenomenological studies.
As discussed above, non of the string limits should be regarded
as fundamental and therefore each limit can at best reveal some
properties of the true nonperturbative vacuum. We should also
consider the possibility that in the end the true fundamental
vacuum may be probed only by its perturbative limits, and that the
underlying nonperturbative theory be defined only for conceptual
consistency. The new understanding of string theory suggests the
following approach. Suppose that we are able to identify in some limit
a class of vacua that appear viable from a phenomenological perspective.
Vacua here refers to specific classes of compactified
manifolds. Further insight into the properties of the phenomenological
vacua may therefore be gleaned by compactifying the other string limits
on the same class of compactified manifolds.

As reasoned above, the essential property of the Standard Particle Model
is its embedding in $SO(10)$ GUT. From this perspective, the primary
guides in the search for phenomenological string vacua should be
the existence of three chiral generations and their $SO(10)$ embedding.
The class of string vacua that we seek are those for which there
exist a limit that preserves the $SO(10)$ embedding of the Standard Model
spectrum. The only perturbative string limit which enables the $SO(10)$
embedding of the Standard Model spectrum is the heterotic $E_8\times E_8$
string. The reason being that this is the only limit that produces 
the spinorial 16 representation in the perturbative massless spectrum.
In this respect it is likely that other M--theory limits provide
more useful means to study other properties of the fundamental vacuum,
such as dilaton and moduli stabilization. 

\section{Perturbative phenomenology}
The study of phenomenological string vacua proceeds with the
compactification of the heterotic string from ten to four dimensions. 
A class string compactifications that preserve the $SO(10)$ embedding of
the Standard Model spectrum are those that are based on the $Z_2\times
Z_2$ orbifold and have been extensively studied by utilizing the
so--called free fermionic formulation\cite{fff,ffm}.
The structure of these models
have been amply reviewed in the past\cite{reviews}.
The models are constructed by specifying a set of 
boundary condition basis vectors and the one-loop GSO projection
coefficients. The first five basis vectors of the realistic free fermionic
models consist of the NAHE set\cite{nahe}.
The gauge group after the NAHE set is
$SO(10)\times E_8\times SO(6)^3$ with $N=1$ space--time supersymmetry, 
and 48 spinorial $16$ of $SO(10)$, sixteen from each sector $b_1$,
$b_2$ and $b_3$, which are
the three twisted sectors of the corresponding $Z_2\times Z_2$
orbifold. The $Z_2\times Z_2$ orbifold is special
precisely because of the existence of three twisted sectors,
that naturally yields three generation models, one from each of the 
twisted sectors. The construction proceeds by adding to the
NAHE set three additional boundary condition basis vectors
which break $SO(10)$ to one of its subgroups\cite{ffm}.
At the same time the number of generations is reduced
to three generations. One spinorial of $SO(10)$, decomposed
under the final $SO(10)$ unbroken subgroup, is obtained
from each of the twisted $b_1$, $b_2$ and $b_3$.
Consequently the weak hypercharge, which arises as
the usual combination $U(1)_Y=1/2 U(1)_{B-L}+ U(1)_{T_{3_R}}$,
has the standard $SO(10)$ embedding. The models contain several
electroweak Higgs multiplets and couplings that may yield
qualitatively viable fermion mass textures\cite{fmt}.

In addition to the standard GUT spectrum, the string models also
contain exotic states which arise from the basis vectors
that break the $SO(10)$ symmetry\cite{ccf}. These states
carry either fractional $U(1)_Y$ or $U(1)_{Z^\prime}$ charge.
Such states are generic in superstring models
and impose severe constraints on their validity.
In some cases the exotic fractionally charged
states cannot decouple from the massless
spectrum, and their presence invalidates otherwise
viable models\cite{otherrsm}.
In the NAHE based models the fractionally
charged states always appear in vector--like
representations, and, in general, mass
terms are generated from renormalizable or nonrenormalizable
terms in the superpotential.
The analysis of ref. \cite{cfn} demonstrated the
existence of free fermionic models with solely the
MSSM spectrum in the low energy effective field theory of the
Standard Model charged matter. 
In general, unlike the ``standard'' spectrum, the ``exotic'' spectrum is
highly model dependent.

The free fermionic string models provide the arena for studying many
of issues that pertain to the phenomenology of the Standard Model
and Unification. Many of these issues have been the subject of
past studies, that include\cite{fmt,top,phenoffm}
among others: top quark mass prediction\cite{top}, several 
years prior to the actual observation by the CDF/D0
collaborations\cite{cdfd0}; generations mass hierarchy; CKM mixing;
superstring see--saw mechanism; Gauge coupling
unification; Proton stability; and
supersymmetry breaking and squark degeneracy.

\section{$Z_2\times Z_2$ orbifold correspondence}\label{z2z2orbifold}

The key property of the fermionic models that is exploited
in trying to elevate the analysis of these models to the 
nonperturbative domain of M--theory is the correspondence
with the $Z_2\times Z_2$ orbifold compactification. 
The correspondence of the NAHE-based free fermionic models  
with the orbifold construction is illustrated
by extending the NAHE set, $\{ 1,S,b_1,b_2,b_3\}$, by one additional   
basis vector $\xi_1$\cite{foc}.
With a suitable choice of the GSO projection coefficients the
model possesses an ${\rm SO}(4)^3\times {\rm E}_6\times {\rm U}(1)^2
\times {\rm E}_8$ gauge group
and $N=1$ space-time supersymmetry. The matter fields
include 24 generations in the 27 representation of
${\rm E}_6$, eight from each of the sectors $b_1\oplus b_1+\xi_1$,
$b_2\oplus b_2+\xi_1$ and $b_3\oplus b_3+\xi_1$.
Three additional 27 and $\overline{27}$ pairs are obtained
from the Neveu-Schwarz $\oplus~\xi_1$ sector.

To construct the model in the orbifold formulation one starts
with the compactification on a torus with nontrivial background
fields. The subset of basis vectors, $\{ 1,S,\xi_1,\xi_2\}$,
generates a toroidally-compactified model with $N=4$ space-time
supersymmetry and ${\rm SO}(12)\times {\rm E}_8\times {\rm E}_8$ gauge
group. The same model is obtained in the geometric (bosonic) language
by tuning the background fields to the values corresponding to
the SO(12) lattice. The
metric of the six-dimensional compactified
manifold is then the Cartan matrix of SO(12),
while the antisymmetric tensor is given by $b_{ij}=g_{ij}$ for $i>j$.
When all the radii of the six-dimensional compactified
manifold are fixed at $R_I=\sqrt2$, it is seen that the
left- and right-moving momenta
reproduce the massless root vectors in the lattice of
$SO(12)$\cite{foc}. 
Adding the two basis vectors $b_1$ and $b_2$
corresponds to the ${Z}_2\times {Z}_2$
orbifold model with standard embedding.
Starting from the Narain model with ${\rm SO}(12)\times
{\rm E}_8\times {\rm E}_8$
symmetry, and applying the ${Z}_2\times {Z}_2$
twist on the internal coordinates, reproduces
the spectrum of the free-fermion model
with the basis
$\{ 1,S,\xi_1,\xi_2,b_1,b_2\}$.
The Euler characteristic of this model is 48 with $(h_{11},h_{21})=(27,3)$,
and it is denoted as $X_2$.
The four dimensional gauge symmetry at this stage can
be either $E_6\times U(1)^2\times SO(4)^3\times E_8$, or
$SO(10)\times U(1)^3\times SO(4)^3\times SO(16)$,
depending on the choice of GSO phase $c({\xi_1, \xi_2})=\pm1$.

The $Z_2\times Z_2$ orbifold of the $SO(12)$ lattice, which is 
realized at the free fermionic point in the moduli space,
differs from the ${Z}_2\times {Z}_2$ orbifold on
$T_2^1\times T_2^2\times T_2^3$, which gives $(h_{11},h_{21})=(51,3)$.
In \cite{foc} it was shown that the two models may be connected
by adding a freely acting twist or shift.
Denoting the three complex coordinates of the 
$T^1_2\times T^2_2\times T^3_2$ tori by  
$z_1$, $z_2$ and $z_3$. Acting with the $\{\alpha,\beta\}=Z_2\times Z_2$
twists on this space produces a model with 48 twisted fixed points,
16 from each of the twisted sectors $\alpha$, $\beta$ and $\alpha\cdot\beta$.
The resulting manifold has $(h_{11},h_{21})=(51,3)$, and is denoted
as $X_1$.
The additional freely acting shift
$
\gamma:(z_1,z_2,z_3)\rightarrow(z_1+{\textstyle{1/2}},z_2+
{\textstyle{1/2}},z_3+{\textstyle{1/2}})
$
produces again fixed tori from the three
twisted sectors $\alpha$, $\beta$ and $\alpha\beta$
and does not produce any additional fixed tori.
Under the action of the $\gamma$-shift,
the fixed tori from each twisted sector are paired.
Therefore, $\gamma$ reduces
the total number of fixed tori from the twisted sectors   
by a factor of ${2}$,
yielding $(h_{11},h_{21})=(27,3)$. This model therefore
reproduces the data of the ${Z}_2\times {Z}_2$ orbifold
at the free-fermion point in the Narain moduli space.
The precise form of the shift that reproduces the $SO(12)$
lattice, and hence the $Z_2\times Z_2$ at the free fermionic point
is discussed in ref. \cite{foc}. 
However, all the models that are obtained
from $X_1$ by a freely acting ${Z}_2$-shift have $(h_{11},h_{21})=(27,3)$
and hence are connected by continuous extrapolations.

The connection between $X_1$ and $X_2$
by a freely acting shift has profound consequences.
The result of adding the freely acting shift $\gamma$
is that the new manifold $X_2$, while still admitting
three twisted sectors, is not simply connected and hence
allows the breaking of the SO(10) symmetry by utilizing
the Hosotani--Wilson symmetry breaking mechanism\cite{hosotani}.
Thus, we can regard the utility of the free fermionic machinery
as singling out a specific class of compactified manifolds. 
In this context the freely acting shift has the crucial
function of connecting between the simply connected covering manifold
to the non-simply connected manifold. Precisely such a construction
has been utilized in \cite{donagi} to construct non-perturbative
vacua of heterotic M-theory.
\begin{figure}[t]
\centerline{\epsfxsize 3.0 truein \epsfbox {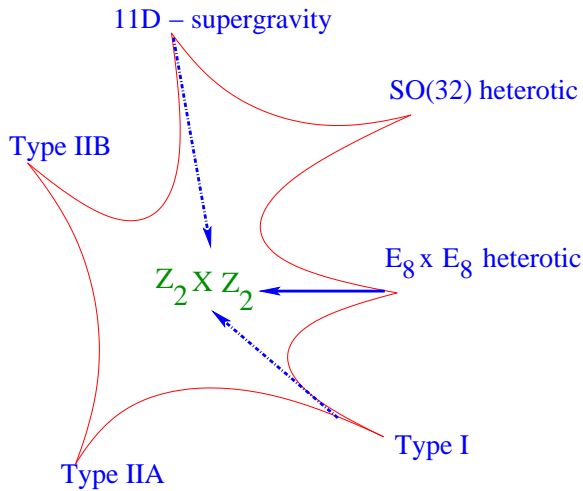}}
\caption{Phenomenological application of M--theory }  
\label{btd02proc1}
\end{figure}
\section{M--embeddings}
The profound new understanding of string theory that   
emerged over the past few years means that we can use
any of the perturbative string limits, as well as eleven
dimensional supergravity to probe the properties of
the fundamental M--theory vacuum.
The pivotal property that this vacuum should preserve
is the $SO(10)$ embedding of the Standard Model spectrum.
This inference follows from the fact that also in the
strong coupling limit heterotic M--theory produces
discrete matter and gauge representations. 
Additionally, the underlying compactification should allow
for the breaking of the $SO(10)$ gauge symmetry.
In string theory the prevalent method to break the
$SO(10)$  gauge group is by utilizing Wilson
line symmetry breaking. Compactification of M--theory 
on manifolds with $SU(5)$ GUT gauge group that can broken
to the Standard Model gauge group were discussed in \cite{donagi}.
In \cite{fgi} the analysis was extended to $SO(10)$ GUT gauge
group that can be broken to $SU(5)\times U(1)$. This
work was reviewed in \cite{reviews} and here I discuss relevant points
for further explorations of the phenomenological free fermionic models.

The key to the construction of ref. \cite{donagi} is the
utilization of elliptically fibered Calabi--Yau threefolds.
These manifolds are represented as a two dimensional
complex base manifold and a one dimensional complex fiber
with a section. 
On these manifolds the equation for the fiber is given 
in the Weierstrass form
$$
y^2=x^3+f(z_1,z_2)x+g(z_1,z_2)=(x-e_1)(x-e_2)(x-e_3).
$$
Here $f$ and $g$ are polynomials of
degrees 8 and 12, respectively and are functions of
the base coordinates; $e_1$, $e_2$ and $e_3$ are the three
roots of the cubic equation. Whenever two of the roots coincide
the fiber degenerates into a sphere. Thus, there is a locus
of singular fibers on the base manifold.
These singularities are resolved by splitting the fiber
into two spherical classes $F$ and $F-N$. One being the
original fiber minus the singular locus, and the second 
being the resolving sphere.

A nonperturbative vacuum state of the heterotic M--GUT--theory 
on the observable sector is specified by a set of 
M--theory 5--branes wrapping a holomorphic 2--cycle on the 3--fold. The 
5--branes are described by a 4--form cohomology class $[W]$
satisfying the anomaly--cancellation condition. 
This class is Poincar\'e--dual 
to an effective cohomology class in $H_2(X, {\bf Z})$ that can be written as
$$
[W]= c_2(TX)-c_2(V_1)-c_2(V_2)=
\sigma_*(w)+c(F-N)+dN,
$$
where $c_2(TX)$, $c_2(V_1)$ and $c_2(V_2)$ are the 
second Chern classes of the tangent bundle and the two
gauge bundles on the fixed planes; $c,d$ are positive definite
integers, $\omega$ is a class in $B$, and
$\sigma_*(\omega)$ is its pushforward to $X$ under $\sigma$.

The key to the M--theory embedding of the free fermionic models
is their correspondence with the $Z_2\times Z_2$ orbifold. The starting
point toward this end is the $X_1$ embedding manifold with 
$(h_{11},h_{21})=(51,3)$. The manifold is then rendered non--simply
connected by the freely acting involution and the methodology of
ref. \cite{donagi,fgi} can be adopted to construct viable M--theory
vacua. The difference however is that now the fiber is
more singular than the ones previously considered.
The fiber of $X_1$ in Weierstrass form is given by
$$
y^2 = x^3 + f_8 (w,\tilde w) x z^4 + g_{12}(w,\tilde w) z^6,
$$
where 
$$
f_8=\eta-3h^2,~{\rm and}~g_{12}~=~h(\eta-2 h^2),
$$
$$
h = K \prod_{i,j=1}^4(w- w_i)(\tilde w - \tilde w_j)
$$ 
and
$$
\eta = C \prod_{i,j=1}^4(w- w_i)^2(\tilde w - \tilde w_j)^2.
$$
Taking $w\rightarrow w_i$ (or ${\tilde w}\rightarrow
{\tilde w}_i$) we have a $D_4$ singular fiber.
These $D_4$ singularities intersect in 16 points, $(w_i,\tilde w_j),\,
i,j=1,\ldots 4$, in the base.
The resolution of the singular fiber in this case
is more involved than the simpler ones previously considered.
It is expected that the richer structure of fiber
classes will yield a richer class of M--theory vacua
with the possibility of new features appearing.

Figure (\ref{btd02proc1}) illustrates qualitatively the approach
to the phenomenological application of M--theory advocated in this
paper. In this view the different perturbative M--theory
limits are used to probe the properties of a specific
class of compactifications. In this respect one may 
regard the free fermionic models as illustrative examples.
Namely, in the heterotic limit this formulation highlighted
the particular class of models that are connected to the
$Z_2\times Z_2$ orbifold. In order to utilize the M--theory
advances to phenomenological purposes, our task then is
to now explore the compactification of the other
perturbative string limits on the same class of
spaces, with the aim of gaining further insight
into their properties. In this spirit compactifications of 
type I string theory on the $Z_2\times Z_2$ orbifold that are
connected to the free fermionic models have been explored\cite{dave}. 
\section{Conclusions}
The Standard Particle Model data suggests two pivotal ingredients
that should be accommodated in the vacuum of the fundamental
quantum gravity theory. The existence of three chiral generations
and their embedding in $SO(10)$ representations. String vacua
based on $Z_2\times Z_2$ orbifold compactifications that admit
these requirements have been constructed. In these construction
the free fermionic point in the moduli space plays an important
role, and may be singled out due to the maximally enhanced symmetries
generated at this point and its relation to the self--dual point
under the T--duality. To go beyond the perturbative analysis
the advances in M--theory has to be employed, that perhaps
will tell us what is special about the $Z_2$ orbifold? In the
heterotic limits of M--theory, the prevailing method to
break the GUT gauge group is the Hosotani--Wilson symmetry
breaking mechanism. A fascinating aspect of this symmetry breaking
mechanism on topologically non--trivial manifolds is that it
gives rise to ultra--massive meta--stable states that provide
different candidates for explaining the cosmic ray events
beyond the GSK cutoff. Developing the experimental and
phenomenological tools to decipher these events will be
the subject of intense activity in forthcoming years\cite{caffarela}. 

\section*{Acknowledgments}
I would like to thank Carlo Angelantonj, Dave Clements,
Alessandro Caffarela, Claudio Coriano,
Emilian Dudas, Richard Garavuso Jose Isidro, Marco Matone, Sander Nooij
and Michael Pl\"umacher for collaboration and discussions.
Work supported in part by the Royal Society and PPARC.

\end{document}